\documentclass[letterpaper,twocolumn,pre,aps,showpacs,floatfix,superscriptaddress,longbibliography]{revtex4-1}
\usepackage[utf8]{inputenc}
\usepackage{amsmath,  amsthm, amsfonts, mathrsfs}
\usepackage{graphicx}
\usepackage{epsfig}
\usepackage{placeins}
\usepackage{physics}
\usepackage{tensor}
\usepackage[usenames,dvipsnames]{color}
\usepackage[colorlinks,linkcolor=blue,citecolor=blue,urlcolor=blue,breaklinks=true]{hyperref}

\usepackage{xr}
\makeatletter

\begin{document} 

 \title{Modern Concepts of Quantum Equilibration Do Not Rule Out Strange Relaxation Dynamics}

 \author{Lars Knipschild}
 \email{lknipschild@uos.de}
 \affiliation{Department of Physics, University of Osnabr\"uck, D-49069 Osnabr\"uck, Germany}

 \author{Jochen Gemmer}
 \email{jgemmer@uos.de}
 \affiliation{Department of Physics, University of Osnabr\"uck, D-49069 Osnabr\"uck, Germany}

 %---------------------------------------------------------------------------------------
\begin{abstract}
Numerous pivotal concepts have been introduced  to clarify  the puzzle of relaxation and/or equilibration in closed quantum systems. All of these concepts 
rely in some way on specific conditions on Hamiltonians $H$,  observables $A$, and initial states $\rho$  or combinations thereof. 
We numerically demonstrate and analytically argue that there is a multitude of pairs $H,A$ that meet said conditions for equilibration and generate some 
"typical" 
expectation value dynamics which means, $\langle A(t)\rangle \propto f(t)$ approximately holds for the vast majority of all initial states. Remarkably we find that, 
while restrictions on the $f(t)$ exist, they do not at all exclude $f(t)$ that are  rather adverse or ``strange'' regarding thermal relaxation.

\end{abstract}

%-----------------------------------------------------------------------------------------------------------------
%-----------------------------------------------------------------------------------------------------------------

\maketitle

\section{Introduction}
\label{intro}

Since the beginning of the previous century, numerous concepts have been suggested to account for the emergence of irreversible thermodynamics from the 
underlying, reversible unitary quantum dynamics \cite{gogolin_equilibration}.
Despite its  long history, this  is an active field of research, even to date. With the paper at hand we intend to show that, while all these concepts are certainly
cornerstones of our understanding of the origins of the second law, they are not yet 
sufficient to rule out certain types of dynamics of observables that may be perceived a being at odds with the arrow of time and which we address as 
``strange dynamics''. But before embarking on a detailed demonstration of their existence, we first name some valuable approaches to 
equilibration in closed quantum systems which nonetheless fail to exclude said strange dynamics. 
This list is neither intended to be complete nor to serve as a full fledged introduction to the subject, readers well acquainted with its items may skip it.

%One central topic in this field is the process of equilibration.
%This refers to the question, how a microscopically constantly evolving system can macroscopically appear to relax to a stable state.

%The most prominent definitions of equilibration refer to time-averages. A time dependent quantity $a(t)$ is said to equilibrate if it is close to 
%some equilibrium-value for most times during its evolution. This type of equilibration is also called "equilibration on average".

{\em equilibration on average:} Statement on temporal fluctuations of e.g., expectation values. Let the  Hamiltonian $H$ with $H\ket{\epsilon_j} = \epsilon_j\ket{\epsilon_j}$  have a sufficiently low number of equal energy gaps
 $\epsilon_i - \epsilon_j$ (``non-resonance condition''). Let furthermore the ``effective dimension'' $d_\text{eff}:=  (\sum_j \bra{\epsilon_j} \rho  \ket{\epsilon_j}^2)^{-1} $ 
 with $\rho$ being the initial state, be large, i.e.,  $d_\text{eff} \gg 1$. Then excursions of an expectation value $\langle A(t)\rangle $ from some 
 ``equilibrium value'' $A_\text{eq}:= \int_0^T \langle A(t)\rangle \text{d}t/T$ are rare in the sense that 
 $ (\langle A(t)\rangle - A_\text{eq})^2 \ll ||A|| $ for most $t$ from $[0,T]$  \cite{reimann_realistic, short2012, short2009}.

{\em eigenstate thermalization hypothesis (ETH):} Assumption on $H,A$, more precisely on random number-like properties  of matrix elements 
$A_{ij} := \bra{\epsilon_i} A  \ket{\epsilon_j}$, see also Eq. (\ref{ETH}). The applicability of the ETH ensures the above rareness of excursions from equilibrium and 
a  fixed $A_\text{eq}$ for all initial states from some energy shell (``thermalization''). The validity of the ETH is closely related to quantum chaos, cf. 
corresponding item below. \cite{srednicki_eth, Deutsch_ETH}.

{\em typicality:} Statistical statement on properties of pure states.
Let the dimension of some finite Hilbertspace $d$ be large, $d \gg 1$. Let $A$ be an operator on this Hilbertspace with a spectral variance of order unity.
Furthermore $| \psi \rangle$ are pure states that drawn at random according to the unitary invariant Haar-measure from this Hilbert space. Then, for the
overwhelming majority of all $| \psi \rangle$, the expectation values $\langle \psi |A| \psi \rangle$ are very close to $\Tr{A}/d$ \cite{lloyd, reimann07, reimann19, Popescu2006, short2009, Tasaki2016, sugiura13, goldstein06}. In the absence of
of any further information on  $| \psi \rangle$ it is thus rational to assume $\langle \psi |A(t)| \psi \rangle \approx \Tr{A}/d$, independent of $t$.

{\em quantum chaos:} An observable featuring a finite overlap with a conserved quantity cannot relax to an equilibrium value independently of its initial value 
\cite{mazur69}.
Wigner-Dyson-type (rather than Poissonian) statistics of spacings of adjacent energy eigenvalues signal the absence of many nontrivial conserved quantities
(``integrability'') \cite{haake1991quantum}.

{ \it microreversibility:} Principal property of the dynamics of a system that ensures that an operation like flipping of particle momenta indeed 
entails a kind of reversed dynamics. Fluctuation relations, which are often viewed as more detailed formulations of the second law, are routinely based on 
microreversibility \cite{campisi}. While the abstract concept is more encompassing, a version of microreversibility is implemented if $A,H$ are both real 
in a common basis, cf. Eq. (\ref{ETH}).

{\em nonfine-tuned initial states:} Some specific mathematical constructions of initial states $\rho$ obviously allow for implementing dynamics that are at odds 
with regular thermal relaxation. Thus a non-trivial conflict with the arrow of time requires the occurrence of strange relaxation dynamics for nonfine-tuned initial states,
i.e., states that do not require: backwards evolving of non-equilibrium states, flipping of particle momenta or complex conjugating of wave-functions 
$|\psi(t)\rangle \rightarrow |\psi(t)\rangle^*  $ (no ``Loschmidt-operations''), full control over each individual matrix element 
$\bra{\epsilon_i} \rho  \ket{\epsilon_j}$, etc. \cite{campisi}

{\color{black}
This article is organized as follows:
	In section \ref{sec_main_claim} we elaborate on what is meant by "strange dynamics" and formulate our main claim.
	In section \ref{sec_construction} we present numerical examples of "strange dynamics" that back up our main claim
	and explain how the numerical construction of these examples works. 
	The physical relevance of the respective initial states underlying these dynamics is discussed in section \ref{sec_phys_init}.
	In section \ref{sec_diag_init} we analytically argue for the validity of the main claim for (mixed) initial states that commute with the observable.
	These arguments suggest that the main claim is valid in the limit of large systems, which is supported by a numerical finite-size-scaling in section \ref{sec_accuracy}.
	In section \ref{sec_non_diag_init} we generalize the result on certain classes of pure initial states that do not commute with the observable.
}

\section{Notion of strange dynamics and main claim} \label{sec_main_claim}

\textcolor{black}{Prior to stating the main claim of the paper at hand we first establish the notion  of  "strange dynamics": To comply with equilibration on 
average the considered expectation value  $\Tr{ A(t) \rho}:=\langle A(t)\rangle$ must take a fixed value  $A_\text{eq}$ for the vast majority of all instances in time.
It may, however, nevertheless exhibit a behavior which is entirely  unexpected 
in the context of relaxation dynamics. It could, e.g., follow the contour of some skyline before settling to $A_\text{eq}$, cf. Fig \ref{dynamics}.
 Or it could, after 
having seemingly settled to $A_\text{eq}$, spike to a significantly  off-equilibrium value at a time long compared to its initial 
relaxation time but much shorter than the Poincare recurrence time, cf. Fig \ref{recurrence}. Any such unexpected  dynamics we call
"strange dynamics" without further rigorous definition. }

Our main claim is as follows: It is possible to find a multitude of pairs $H,A$ in accord with all above equilibration principles such that
\begin{equation}
\label{newmain} 
 \langle A(t) \rangle \approx  \langle A(0) \rangle f(t) \quad (f(0)=1)
\end{equation}
\textcolor{black}{where $f(t)$ must have a positive Fourier transform. Other than that $f(t)$ may essentially be freely chosen. Among the possible choices are plenty of 
strange dynamics, cf. Figs. \ref{dynamics}, \ref{recurrence}}.
Most importantly the validity of 
Eq. (\ref{newmain}) is claimed \textcolor{black}{for a vast majority of all initial states $\rho$. This major set of initial states will be detailed below, see Sects. 
\ref{valcom}, \ref{valnocom}}.

This claim is the more technical reformulation of the central statement which forms the title of the present paper. (For some numerical evidence of the validity of 
Eq. (\ref{newmain}) in concrete spin systems, albeit with non-strange $f(t)$'s, see Ref. \cite{Richterprinzip_2018})

\section{Making of Figures 1 and 2: Hamiltonians, Observables and Initial States} \label{sec_construction}

\begin{figure}
	\includegraphics[width=0.48\textwidth]{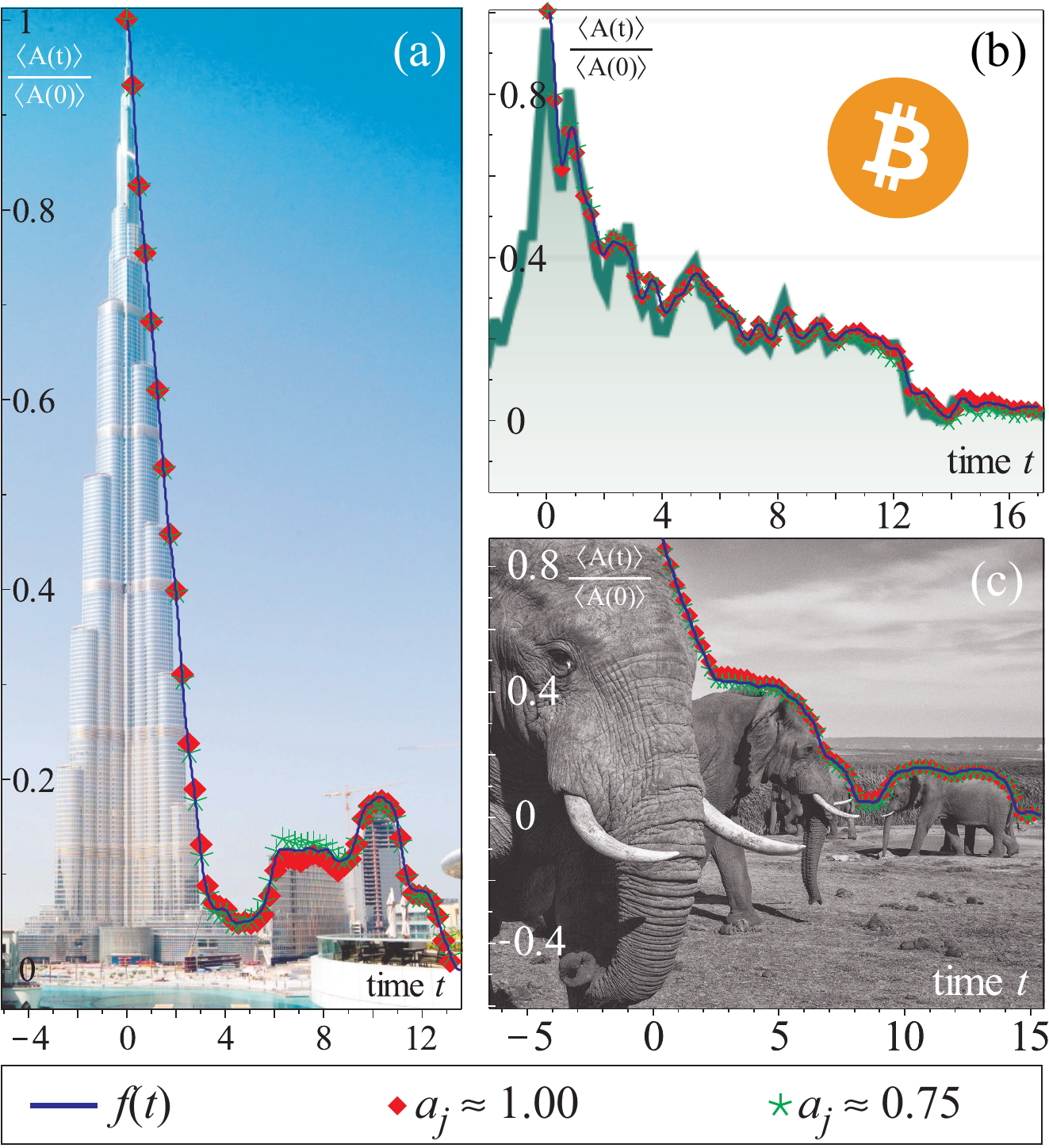}
	\caption{\textcolor{black}{$H$ and $A$ may be tailored such that $\langle A(t) \rangle/ \langle A(0) \rangle$ follows almost any desired function $f(t)$ 
	for a majority of initial states. Displayed data are for initial states that are eigenstates of $A$, i.e.,  
	$A|\psi(0)\rangle = a_j|\psi(0)\rangle $. Green and red symbols: $\langle A(t) \rangle/ \langle A(0) \rangle$, blue line: desired function $f(t)$. } \\
	\textbf{(a)} Burj Khalifa (original photo: "Burj Khalifa" by Joi is licensed under CC BY 2.0, https://commons.wikimedia.org), 
	\textbf{(b)} Bitcoin price, 
	\textbf{(c)} elephants (original photo: "Family Of Elephants" by Javier Puig Ochoa is licensed under CC BY 3.0, https://commons.wikimedia.org)
	}
		   \label{dynamics}
	\end{figure}

\begin{figure}
		\includegraphics[width=0.48\textwidth]{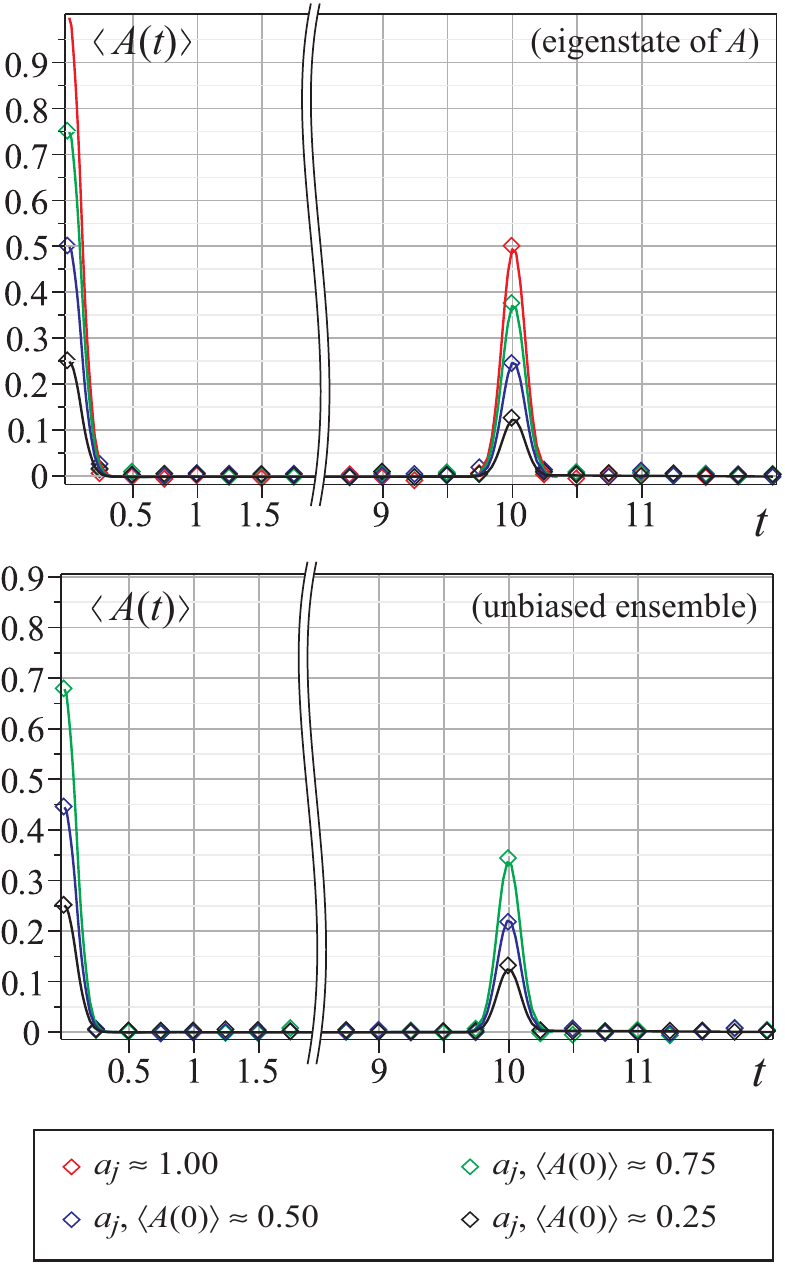} 
		\caption{$\langle A(t) \rangle$ exhibits a significant revival at an arbitrary predefined time $\tau_R$ (here $\tau_R=10$) for two classes of  
		initial states: (a) $|\psi(0)\rangle$ are eigenstates of $A$, (b)  $|\psi(0)\rangle$ are samples of an unbiased pure state ensemble conditioned 
		on a given  $\langle A(0) \rangle$, cf. Eq. (\ref{unbiased})}
			\label{recurrence}
		\end{figure}

\textcolor{black}{Figs.  \ref{dynamics}, \ref{recurrence} show various expectation value dynamics as resulting from the solution of the Schr\"odinger equation for 
fixed $H, A$ (per panel in  Fig.  \ref{dynamics})     but various $\rho$. The somewhat odd examples in Fig.  \ref{dynamics} have simply been picked  to substantiate 
the above claim that $f(t)$ may essentailly be chosen at will,
cf. Eq. (\ref{newmain}). The setting in  Fig. \ref{recurrence} is meant  as a prime example of strange dynamics in the sense described above.}
Of course peculiar  dynamics as presented in Figs.  \ref{dynamics}, \ref{recurrence} require pairs $H,A$ with specific properties. In the following we
 detail the construction of $H,A$, thereby unveiling their accordance with the cornerstone principles of  thermal relaxation.

First a $d$-dimensional Hamiltonian $H$ is defined by choosing  $d$ eigenvalues $\epsilon_j$ (examples  in Figs. \ref{dynamics}, \ref{recurrence}: $d=20000$).
To this end $d-1$ energy-gaps $l_j = \epsilon_{j+1} - \epsilon_{j}$ are drawn as i.i.d. random numbers from a pertinent  Wigner-Dyson distribution. 
The spectrum is scaled to span an interval $[-E, E]$ (examples  in Figs. \ref{dynamics}, \ref{recurrence}:  $E=30$).
%\begin{equation}
%\label{luft}
%E \gg  \omega_\text{max}
%\end{equation}
%(below examples  in the present paper:  $E=30, \omega_\text{max} \approx 3 $). 
Within this interval $H$ has thus a constant density of states and 
exhibits Wigner-Dyson 
level statistics as expected for non-integrable systems. Thus our modeling is in accord with the non-resonance condition and quantum chaos in the sense of the respective 
items in the introductory list in Sect. \ref{intro}.
This  Hamiltonian $H$ may be viewed as the sector of a, e.g., many-body Hamiltonian that corresponds to a (narrow) energy window stretching from $-E$ to $E$.
If the initial state lives (almost) entirely in this energy window, which is what we assume here, modeling of this sector suffices to compute the dynamics.
\textcolor{black}{Note that $\langle A(t) \rangle$ is fully  determined by the spectra and the "relative angles" of the eigenvectors of the operators $H, A, \rho$. Thus, 
there is 
no need to specify the eigenvectors of $H$ with respect to some "computational basis" for the purposes at hand.}

Next we construct the observables $A$. To this end we first define  $\tilde{f}(\omega)$ to be the real part of the Fourier transform of some 
desired, possibly strange,  $f(t)$. 
Let furthermore  $\omega_\text{max}$ be some 
``cut-off frequency''  $\omega_\text{max}$, such that  $\tilde{f}(\omega)$  
attains only negligible values at  $|\omega| \geq \omega_\text{max}$. Choose $E,\omega_\text{max}$ such as to fulfil $ E \gg \omega_\text{max}$ 
(examples  in Figs. \ref{dynamics}, \ref{recurrence}: $\omega_\text{max} \approx 3 $).    
Furthermore $\tilde{f}(\omega)$ has to vary only negligibly on the scale of
the level-spacings $l_j$. While these conditions on $\tilde{f}(\omega)$ imply conditions on  $f(t)$, these conditions become exeedingly mild,
at sufficiently large $d$.

Now we specify the  $A$'s in the energy eigenbasis   $\{\ket{\epsilon_i}\}$ in full accord with the ETH \cite{srednicki99} as
\begin{equation}
\label{ETH} 
  {A}_{jl}  :=   \bra{ \epsilon_j}A\ket{\epsilon_l} = C_1 d^{-1/2}\sqrt{\tilde{f}(\epsilon_j - \epsilon_l   )} \, R_{jl} \ ,
\end{equation}
where  $R_{jl}$ are normally  i.i.d. random real numbers  with zero mean and unit variance. $C_1$ is a constant which we use to scale the 
extreme eigenvalues of $A$ to $|a_\text{min}| \approx |a_\text{max}|\approx 1$. To render $A$ Hermitian,  $\tilde{f}(\omega)$ must be nonnegative which implies 
the condition on $f(t)$ mentioned below Eq (\ref{newmain}). \textcolor{black}{To support our main claim it suffices  that 
$A$ as defined in Eq. (\ref{ETH}) is not in conflict with the ETH. But what is more, numerous numerical studies found operators of local observables 
(currents, magnetizations, etc.) in the eigenbasis' of the respective Hamiltonians (many-body lattice models) to essentially  agree with the 
construction  Eq. (\ref{ETH}) 
\cite{Rigol_ChaosETH_2016, Haque_2015, Rigol_ETH_2017, mukerjee06, kolovsky04, wimberger14, torres-herrera16,  nation18, mondaini18, hamazaki19, foini19a}, 
for more details see App. \ref{eth_physical}.} Choosing the ${A}_{jl}$ real renders the set-up  microreversible in the sense defined in the respective item of the 
introductory list  of equilibration principles .

Ultimately we aim at establishing that this construction of $H,A$ renders Eq. (\ref{newmain}) valid for 
practically all nonfine-tuned initial states $\rho$. However, for the numerical demonstration of this validity 
 as displayed in Figs.  \ref{dynamics}, \ref{recurrence}, we chose initial states from the following two 
classes:

{\em observable eigenstates:} Let $| a_j \rangle$ be an eigenstate of $A$, i.e., $A | a_j \rangle = a_j| a_j \rangle$. 
The dynamics in Figs. \ref{dynamics}, \ref{recurrence} 
(upper panel) are computed for some  sample initial eigenstates of $A$, i.e.,  $\rho = | a_j \rangle \langle a_j |$ for eigenvalues $a_j$ 
which may be inferred from the 
respective captions.

\textcolor{black}{{\em unbiased ensemble states:}
The dynamics in 
Fig. \ref{recurrence} (lower panel) correspond to states   $\rho = | \psi \rangle \langle \psi |$  with
\begin{equation}
\ket{\psi} =  \bra{\phi}\ket{\phi}^{-1/2}   \ket{\phi}  , \quad 
\ket{\phi} = (1 + \lambda A)^{-1/2}\ket{\xi}
\label{unbiased}
\end{equation}
where $\ket{\xi}$ is a random vector unitarily drawn from the $d$-dimensional hypersphere in Hilbertspace. $\lambda$ has been chosen such as to obtain 
the respective $\langle A(0) \rangle $ as  indicated in the caption of Fig. \ref{recurrence}.}

\textcolor{black}{Before motivating the above choice of initial states we provide some more information on Figs. \ref{dynamics}, \ref{recurrence}.
For Fig. \ref{dynamics} the 
functions $f(t)$ have 
been extracted from the respective underlying pictures. 
(The pictures and contours have been chosen such as to make sure that the contours indeed feature positive Fourier transforms.)  
For Fig. \ref{recurrence} $f(t)$ is simply a pertinent mathematical function, 
the revival time $\tau_R$ of which is chosen as $\tau_R=10$, which is much longer than the timescale of the initial decay but definitely much shorter than
the Poincare recurrence time, cf. \ref{decay_fidelity}, making this example a prime instance of strange dynamics.}
 The respective $H, A$ have been constructed according to the above scheme. 
The respective Schr\"odinger equations have been numerically solved for the sample initial states as described above and in the captions.
Obviously all computed data are in excellent agreement with our main claim, Eq. (\ref{newmain}). (Even more numerical support for the latter comes from the data 
displayed in Fig \ref{error_dist}). All initial states feature effective dimensions of at least $d_\text{eff} \geq 4700$, i.e., are in accord with the 
principle of equilibration on average, for more detailed information see App. \ref{app_effective_dimensions}.

\section{Physical Significance of the initial states underlying the displayed data} \label{sec_phys_init}

Before explaining the inner workings of the above construction of $H,A$ we comment on the physical significance of the initial states used  
in Figs. \ref{dynamics}, \ref{recurrence}. \textcolor{black}{ Some physical relevance of the initial observable eigenstates $| a_j \rangle$ comes from their being 
identical to results   
		of (preparatory) projective measurements of the monitored observable $A$ at the beginning 
		of the relaxation dynamics. More importantly, however, the validity of  Eq. (\ref{newmain}) for $\rho = | a_j \rangle \langle a_j |$ for all $j$ 
		necessarily   entails
		the validity of Eq. (\ref{newmain}) also for all initial states of the form $\rho =\sum_n c_n A^n$.
		The latter comprises e.g. $\rho \propto \exp^{\lambda A}$,
		 which is the state of maximum von-Neumann entropy conditioned on some given $\langle A(0) \rangle $.
		This is a relevant class of initial states within the framework of Jayne's principle.}
		
The states from the unbiased ensemble, cf. Eq. (\ref{unbiased})		
 represent an ensemble of pure states which is entirely unbiased with respect to the unitary invariant Haar-measure, 
under the condition of a given $\langle A(0) \rangle $  \cite{fine09,mueller2011,reimanngemmer}. This ensemble thus is a very 
relevant class of initial states within the framework of pure state statistical mechanics. 

\section{Validity of the main claim (Equation (1)) for initial states which commute with the observable} \label{sec_diag_init}
\label{valcom}

In the following  we explain why the above construction yields pairs $H, A$ that render our main claim Eq. (\ref{newmain}) valid for the above mentioned class of 
initial states $\rho =\sum_n c_n A^n$ that also comprises the observable eigenstates $| a_j \rangle \langle a_j|$. 
Consider the relation 
	   \begin{equation}
	   \Tr{{A}(t) {A}^N } \approx \left \{
	   \begin{array}{ll}
	   \propto \Tr{{A}(t) {A}} \propto f(t) \, , & \text{odd } N \\
	   0 \, , & \text{even } N
	   \end{array}
	   \right. \, . \label{assumption}
	   \end{equation}
	   Obviously the validity of this relation entails the validity of Eq. (\ref{newmain})
	   for all initial states of the above class. Accordingly Eq. (\ref{assumption}) is  the first pillar on which Eq. (\ref{newmain}) rests. 
	   The validity of a subpart of Eq. (\ref{assumption}), namely $\Tr{{A}(t) {A}} \propto f(t)$ follows rather 
	   straightforward from Eq. (\ref{ETH}) (for 
 details see App. \ref{auto_corr_explicit}). To fully validate Eq. (\ref{assumption}) we employ a scheme suggested in Ref. \cite{Richterprinzip_2018}.
 As the argument is quite involved for large exponents 
	   $N$, we here restrict ourselves to $N = 2$ and $N =3$. A comparable but more complex derivation for arbitrary but fixed $N$ at large $d$
	   can be found in App. \ref{appendix_richter}.
	   % More precisely, 
	   % we will use the of-diagonal part of the ETH ansatz, i.e., the second term in 
	   % Eq.\ (\ref{ETH}). 

	   We start by writing out the correlation function for $N = 2$ explicitly, 
	   \begin{equation}
	   \Tr{ {A}(t) {A}^2 } = \sum_{a,b,c} {A}_{ab} A_{bc} {A}_{ca} \, e^{\imath (\epsilon_b-\epsilon_a) t}\ .
	   \label{start}
	   \end{equation}
	   Given the matrix structure in Eq.\ (\ref{ETH}), the biggest part of the addends 
	   in the sum are by construction (products of) independent random numbers with 
	   zero mean.  Thus, to an accuracy set by the law of large numbers, summing the 
	   latter yields zero as well.  There are, however, index combinations for which not 
	   all factors within the addends have vanishing mean, namely, $c = a$. Focusing on 
	   these terms, we can write
\begin{equation}
	   \Tr{ {A}(t) {A}^2 } \approx \!\! \sum_{a,b} |{A}_{ab}|^2 {A}_{aa} e^{\imath (\epsilon_b-\epsilon_a) t}    \, . \label{FC2}
	   \end{equation}
	   While the numbers $|A_{ab}|^2$ do not have mean zero, the numbers $A_{aa}$ do have zero mean. Furthermore 
	   for $a \neq b$ the $|A_{ab}|^2, {A}_{aa}, e^{\imath (\epsilon_b-\epsilon_a) t}$ are mutually independent stochastic variables, cf. Eq.\ \eqref{ETH}. For 
	   $a = b$ these numbers are obviously not independent, however, in this case the sum in Eq.\ \eqref{FC2} 
	   is proportional to the third moment of the distribution of the  $A_{aa}$ which vanishes according to  
	   Eq.\ \eqref{ETH}. Exploiting these findings for both cases ($a\neq b$ as well as $a=b$) to evaluate Eq.\ \eqref{FC2} we obtain $\Tr{{A}(t) {A}^2 } \approx 0$, i.e., 
	   Eq.\ (\ref{assumption}) for the 
	   even case $N = 2$.

	   Now we turn to $N=3$, i.e.,  
	   \begin{equation}
	   \Tr{{A}(t) {A}^3}= \!\! 
	   \sum_{a,b, c,d} {A}_{ab} {A}_{bc} {A}_{cd} {A}_{da} e^{\imath (\epsilon_b-\epsilon_a) t} \, .
	   \end{equation}
	   Again, the contributions of most addends approximately cancel each other upon 
	   summation.  But also here, there are exceptions, namely, the index combinations 
	   $c = a$ or $d = b$. Focusing on these terms, we find
	   \begin{equation}
	   \Tr{{A}(t) {A}^3}\approx \!\! \sum_{a,b} 
	   (|{A}_{ab}|^2 \sum_c |{A}_{bc}|^2 + |{A}_{ac}|^2)  e^{\imath (\epsilon_b-\epsilon_a) t}     \, . \label{FC4}
	   \end{equation}
	   (Note that Eq. (\ref{FC4}) erroneously  counts the terms corresponding to $c = a$ and $d = b$ twice. 
	   However, as this over-counting error is of order $d^{-1}$, it becomes negligible at large $d$.)
	   To proceed, consider the 
	   above sums over $c$ first. 
	   %without the diagonal elements i.e., $\sum_{c \neq b} |{A}_{bc}|^2+\sum_{c \neq a}|{A}_{ac}|^2$.
	   While these sums do not vanish, they are practically independent of $a,b$, due to the specific matrix structure of $A$, cf. Eq.\ \eqref{ETH}.
	   Thus, the respective sums may be replaced by a 
	   constant $C_2$, i.e., $\sum_c |A_{bc}|^2+|{A}_{ac}|^2 \approx C_2$. 
		Inserting this into
		Eq.\ (\ref{FC4}) yields 
		$\Tr{{A}(t) {A}^3}\approx C_2 \cdot  \sum_{a,b} |{A}_{ab}|^2  e^{\imath (\epsilon_b-\epsilon_a) t}   $.
		Comparing this to the exact relation $\Tr{ {A}(t) {A} }=\sum_{a,b}  |{A}_{ab}|^2  e^{\imath (\epsilon_b-\epsilon_a) t} $ yields
		$\Tr{ {A}(t) {A}^3 } \approx C_2\, \Tr{ {A}(t) A}$, i.e., Eq.\ (\ref{assumption}) for the odd case $N = 3$.

\section{Accuracy  of the main claim (Equation (1)) for eigenstates of the observable as initial states}  \label{sec_accuracy}
	   
\begin{figure}
\includegraphics[width=0.48\textwidth]{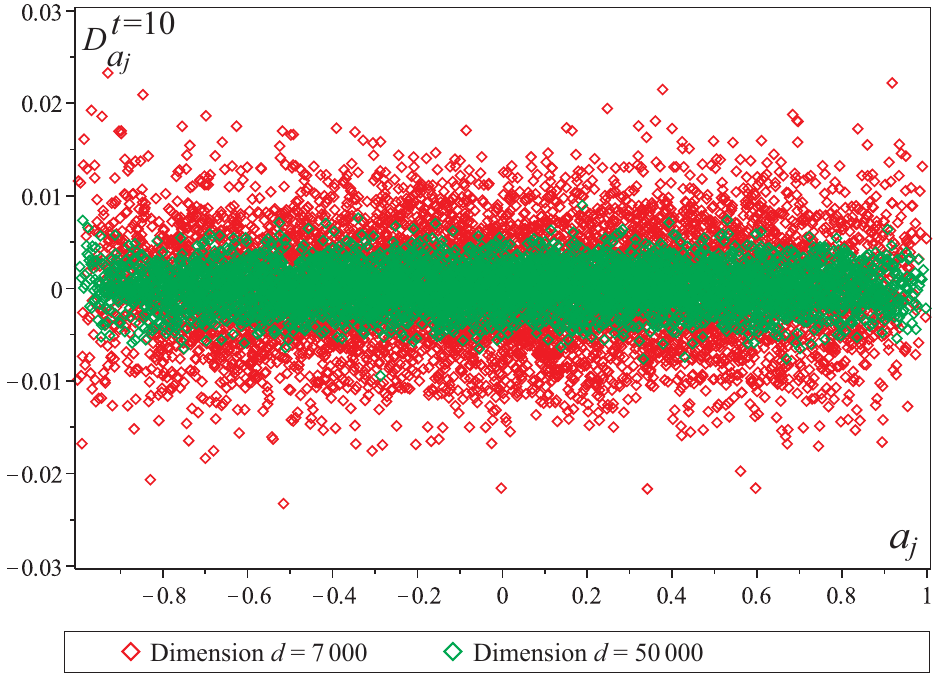} 
\caption{ The mean values of the errors and the variances are almost independent of the eigenstate, which served as the respective initial state.
}
	\label{error_dist}
\end{figure}

	  Mainly due  to the neglect of terms with random signs (cf. below Eq. \ref{start}), Eq. (\ref{assumption}) is not exact. To scrutinize the accuracy of   
	   Eq. (\ref{assumption}) and hence Eq. (\ref{newmain}) specifically with respect to growing dimensions $d$, 
	   the Schr\"odinger 
	   equation has been solved for the set-up underlying Fig. \ref{dynamics}, panel (a), for  all $| a_j \rangle$ (d =7000) respectively  every 
	   10th 
	   $| a_j \rangle$ (d=50 000) as initial states. Figure \ref{error_dist} shows the deviations from Eq. (\ref{newmain}) 
	   at an exemplary point in time, here chosen as $t=10$. 
	   These deviations are defined as 
\begin{equation}
	   D_{a_j}^{t} := \langle a_j |A(t) | a_j \rangle - a_jf(t) .
\end{equation}

The errors have vanishing mean and a standard deviation that is almost independent of $a_j$. But, 
most importantly,  the standard  deviation decreases with the dimension $d$.

To numerically analyze this dependence, we consider the mean square of these errors (averaged over all eigenstates $a_j$ and over time):
\begin{equation}
 \tensor*[]{D}{_d} = \frac{1}{d} \sum_{j=1}^d \frac{1}{T} \int_0^T \left( \tensor*[]{D}{^{t}_{a_j}} \right) ^2 \mathrm{d}t, \quad T=13
\end{equation}

\begin{figure}
\centering
\includegraphics[width=0.48\textwidth]{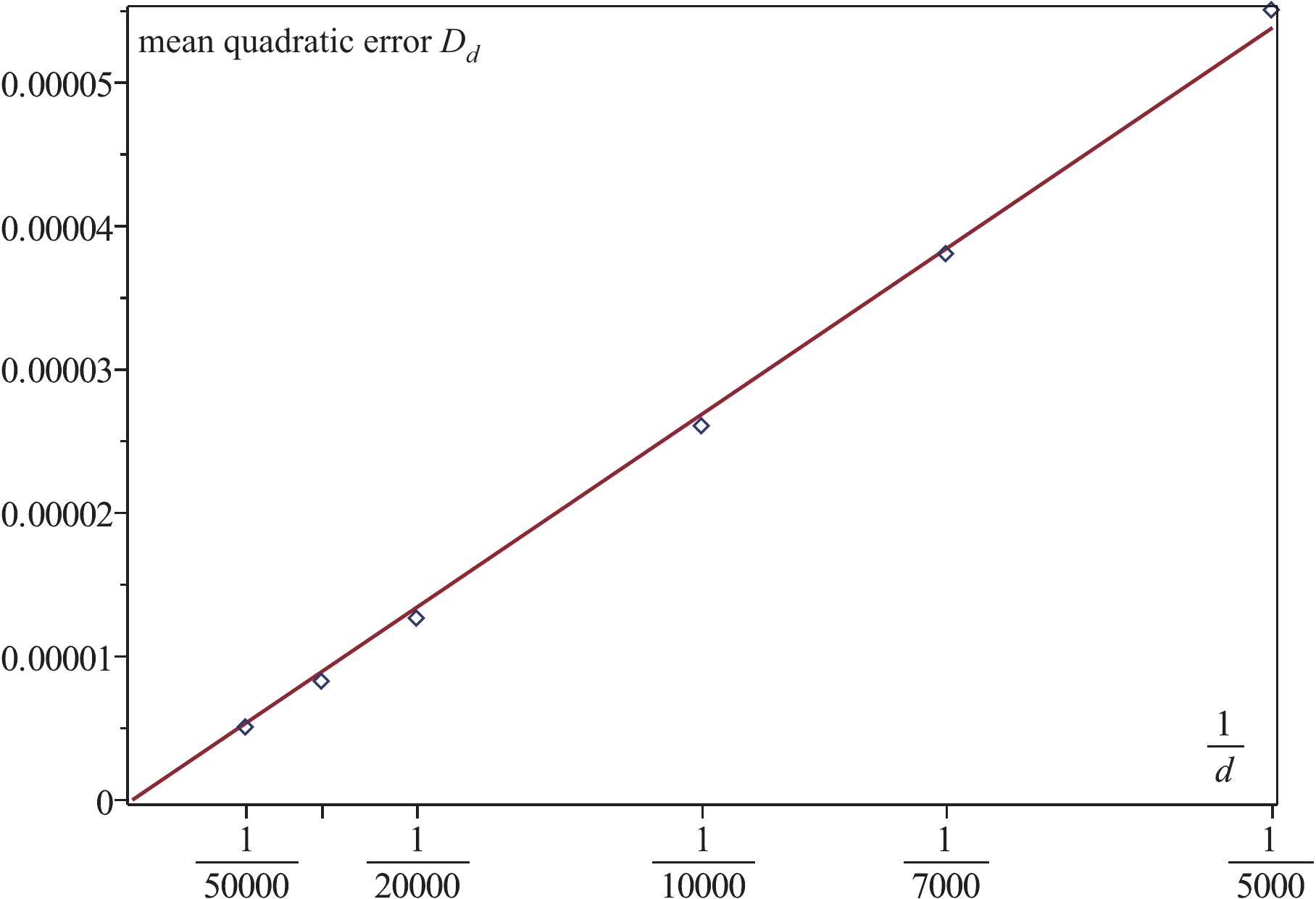} 
\caption{
	The finite-size scaling of the mean square of the errors $D_d$ indicates that these are proportional to $\frac{1}{d}$.
}
	\label{Richter_Errors_finite_size}
\end{figure}

Plotting $D_d$ over the reciprocal Hilbertspace dimension $1/d$ (see Fig. \ref{Richter_Errors_finite_size}) reveals that the mean
square of these errors vanishes as $\propto d^{-1}$. 
This is supported by an analytical reasoning (see App. \ref{appendix_richter}). These findings imply that Eq. (\ref{newmain}) becomes exact in the limit of large 
$d$ for all  initial states of the form $\rho =\sum_n c_n A^n$, i.e., all initial states that commute with the observable $A$.

\section{Validity of the main claim (Equation (1)) for  a class of initial states which do not commute with the observable} \label{sec_non_diag_init}
\label{valnocom}
	 
	 \textcolor{black}{ Next we establish the validity of Eq. (\ref{newmain}) for a set of (pure) initial states that are not of the form $\rho =\sum_n c_n A^n$, i.e., not functions
	   of the observable $A$, and comprises 
	   the unbiased ensemble states (cf. Eq. (\ref{unbiased})). We call this set the ``$r$-set'' and define its  states $\ket{\eta}$ by
	   \begin{equation}
	   \ket{\eta} =\bra{\chi}\ket{\chi}^{-1/2} \ket{\chi}, \quad 
	   \ket{\chi} = \sqrt{r(A)}\ket{\xi}
	   \label{offdia}
	   \end{equation}
	   where $r(A)$ is a nonnegative function that varies little on the scale of the  eigenvalue spacings of  $A$ and $\ket{\xi}$ is a random vector
	    drawn from the
	   unitary invariant Haar-measure.} Typicality arguments may be used to show that for all $\ket{\eta}$,
	   except for a fraction of at most $\propto d^{-1/2}$ the approximation 
	   \begin{equation}
		\label{approx}
	   \bra{\eta} A(t) \ket{\eta} \approx \Tr{r(A)  A(t)}\Tr{r(A)}^{-1}
	   \end{equation}
	   holds to very good accuracy. For a complete derivation see, \cite{reimanngemmer} and App. \ref{app_typicality}. As $r(A)$ may be cast into the form $r(A) =\sum_n b_n A^n$, 
	    the combination of Eqs. (\ref{approx}) and (\ref{assumption}) establishes the validity of Eq. (\ref{newmain}) for all $\ket{\eta}$ except for the above fraction
		of size  $\propto d^{-1/2}$.  Thus Eq. ((\ref{approx}) is the second pillar on which Eq. (\ref{newmain}) rests.   
		For the special case  $r(A)=(1+\lambda A)^{-1}$ the $r$-set is identical to the 
	    unbiased ensemble, cf. Eq. (\ref{unbiased}), hence Eq. (\ref{approx}) explains the numerical findings displayed in
	    Fig. \ref{recurrence} (lower panel). \textcolor{black}{ Note that the $r$-sets are very encompassing. 
		The small fraction of order $d^{-1/2}$ of initial states that does not comply with Eq. (\ref{approx}) for a suitable $r(A)$
		thus corresponds
	    to the set of fine 
	    tuned initial states which always exist but are excluded from the analysis at hand, c.f. last item of the introductory list of principles of equilibration.
	    Note furthermore that the validity of 
	    Eq. (\ref{approx}) entails the  validity of Eq. (\ref{newmain}) for all initial states of the form 
	    $\rho = \int g(\eta) \ket{\eta}\bra{\eta} $d$\eta$ with  $g(\eta)\geq 0$ for the few (fine tuned) $\eta$ to which Eq. (\ref{approx}) does not apply. 
	    Thus Eq. (\ref{newmain}) also holds for a large set of mixed states.}

\section{Outlook}

	While the body of this paper aims at pointing out an overlooked loophole in the current approach to equilibration in closed quantum systems, we eventually very
	briefly turn to possible closings of this loophole: While pairs $H,A$ giving rise to strange dynamics definitely exist, these dynamics may not be stable under 
	 perturbations of the respective Hamiltonians \cite{dabelow19, stability, richter19a}. Apart from that the ETH may miss some correlations  that are in fact present 
	 in the matrices representing physical observables in physical systems \cite{foini19}. These correlations may possibly rule out strange dynamics. 
	 Furthermore arguments are viable that are based on the condition of Hamiltonians being local \cite{kastner19}. Helpful insights may also come from 
	  clarifying the role of Markovianity in closed quantum systems 
		\cite{figueroa-romero19}.

	   {\it Acknowledgements:} 
	   Stimulating discussions with M. Srednicki, K. Modi, M. Rigol and W. Zurek during the QTHERMO18 program at the KITP are gratefully acknowledged. The authors also benefited
	   from interacting with R. Steinigeweg, J. Richter and R. Heveling.
	   This work has been funded by the Deutsche
	   Forschungsgemeinschaft (DFG) - Grants No. 397107022
	   (GE 1657/3-1), No.
	   355031190 - within the DFG Research Unit FOR 2692.
	   Furthermore this research was supported in part by the National Science Foundation under Grant No. NSF PHY-1748958.
	   
\bibliography{mybib}{}

\appendix

\section{Eigenstate Thermalization Hyopthesis in Physical Models} \label{eth_physical}
The similarity of local operators represented in the energy eigenbasis with random matrices, as implemented in Eq.  (\ref{ETH}), is at the heart of the 
ETH \cite{srednicki99}. Such a 
similarity has numerically  been observed frequently, see, e.g., Refs. 
\cite{Rigol_ChaosETH_2016, Haque_2015, Rigol_ETH_2017, mukerjee06, kolovsky04, torres-herrera16,  nation18, mondaini18, hamazaki19, foini19a}, 
it may, however, require a splitting of $A$ into distinct symmetry-sectors. The more specific form of $A$, namely the dependence of the 
``envelope-function'' $\tilde{f}(x)$ solely through the energy differences $\epsilon_j - \epsilon_l$ but not on individual energies, has approximately also been found 
for 
local observables in interacting lattice-particle models, see Refs. \cite{Richterprinzip_2018, jansen19, mukerjee06}. Currently discussed structural differences of 
local observables in physical chaotic systems from  the ETH as formulated in \cite{srednicki99} include non-gaussian distributions of the matrix elements \cite{luitz16} and 
correlations between individual
matrix elements \cite{foini19}. While non-systematic checking indicates that non-gaussian distributions  in Eq. (\ref{ETH}) leave the validity of Eq. (\ref{assumption}) 
unaltered, the impact of correlations is open and subject to further research. However, some evidence for the applicability of the overall concept discussed in the 
paper at hand to currents in spin-chains comes from Ref. \cite{Richterprinzip_2018}.

\section{Decay of Fidelity} \label{decay_fidelity}
In this Section we check the overlap of the evolved  state with the initial state, to exclude that the revival peaks in Fig. \ref{recurrence}
are due to  Poincare-recurrences.
We therefore calculate the fidelity $F$ of the time-evolved state $U(t) \ket{a_j}$ and the respective initial state $\ket{a_j}$:
\begin{equation}
F(t) = |\bra{\psi_0} U(t) \ket{\psi_0}|^2
\end{equation}
$U(t)$ denotes the time-evolution operator for the time $t$.
\begin{figure}
\centering
  \includegraphics[width=0.48\textwidth]{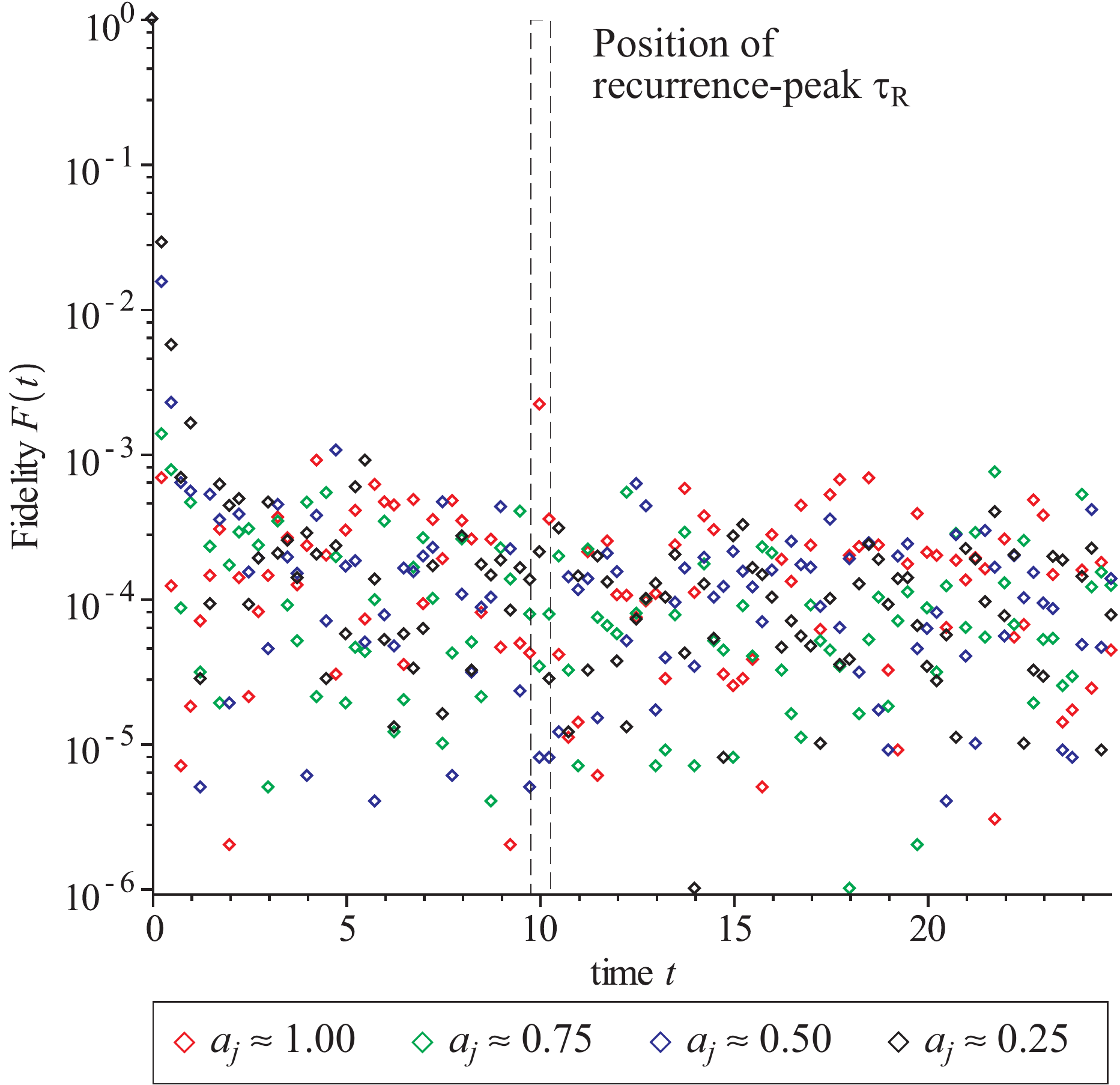} 
\caption{ For the dynamics of the eigenstates, corresponding to the eigenvalues $a_j$, the fidelity has been plotted over time.
	The Fidelity rapidly decays and does not show any significant recurrences. Especially at the position of the recurrence-peak $\tau_R$ (of the expectation-value dynamics)
		the fidelity is less than $0.01$.
}
	\label{fidelity}
\end{figure}
The fidelity rapidly decays  during the initial relaxation of the expectation value. There is no significant increase of $F(t)$ at any later time, including the 
revival time of the expectation value  $\tau_R$.
Thus the second peak in the expectation-value dynamics (at time $\tau_R$) is not due to any Poincare-recurrence.

\section{Effective Dimension of Various Initial States} \label{app_effective_dimensions}
A sufficiently large effective dimension of the initial state is a necessary condition, when trying to prove the emergence of thermodynamic behaviour
in closed quantum systems on the basis of equilibration on average (see Introduction).
Table \ref{effective_dimensions} lists the effective dimensions $d_\mathrm{eff}(\ket{\psi})$ of various  initial states that are either eigenstates $\ket{a_j}$ of $A$
or members of the unbiased ensemble $\ket{\psi_a}$, Eq. (\ref{unbiased}). For the latter the parameter $\lambda$ has been chosen such, that the
expectation value of $A$ for the ensemble average is equal to $\langle A \rangle = 0.25, 0.50, 0.70$.

 Obviously, $d_\mathrm{eff}(\ket{\psi})$ is rather large at all instances.

\begin{table}
\caption{We calculated the effective dimension for various initial states. The data refers to the recurrence-dynamics (\ref{recurrence}).}
\centering
\begin{tabular}{ll}
\hline
\textbf{Initial State} & \textbf{Effective Dimension $d_\mathrm{eff}$} \\
\hline
$\ket{a_j}, a_j=0.25$ & 6500 \\
$\ket{a_j}, a_j=0.50$ & 6500 \\
$\ket{a_j}, a_j=0.75$ & 6400 \\
$\ket{a_j}, a_j=1.00$ & 4700 \\
\hline
$\ket{\psi_{a}}, \langle A(0) \rangle = 0.25$ & 9900 \\
$\ket{\psi_{a}}, \langle A(0) \rangle = 0.50$ & 8900 \\
$\ket{\psi_{a}}, \langle A(0) \rangle = 0.75$ & 7500 \\
\hline 
\label{effective_dimensions}
\end{tabular}
\end{table}

\section{Explicit Calculation of the Auto-Correlation Function} \label{auto_corr_explicit}
Within this Section we calculate the auto-correlation function that follows from Eq.(\ref{ETH}):
\begin{eqnarray} 
\Tr{A(t) A} &=& C_1^2 \sum_{j,l} \left|a_{jl}\right|^2 \cos( (\epsilon_l-\epsilon_j)    t) \label{autocorr1}\\
&\approx&  C_1^2  \sum_{j,l}\tilde{f}(\epsilon_l-\epsilon_j) \cos((\epsilon_l-\epsilon_j)    t) \label{autocorr2} \\
&\approx& C_3  \int \tilde{f}(\omega) \cos(\omega   t) \text{d} \omega \label{autocorr3}\\
& \propto & f(t) \quad  \text{for} \quad t\geq 0 \label{autocorr4}
\end{eqnarray}
Eq. (\ref{autocorr2}) follows from the law of large numbers and for Eq. (\ref{autocorr3}) we exploit the uniform density of states of $H$ and $E \gg \omega_\mathrm{max}$. $C_3$ is a 
pertinent constant. In Eq. (\ref{autocorr4}) we used the definition of the positive Fourier transform. Hence this construction implemented by Eq.(\ref{ETH}) 
essentially produces a autocorrelation function following predefined 
target-dynamics $f(t)$ while being in accord with the ETH.

\section{Analysis of the validity of Equation (\ref{assumption})} \label{appendix_richter}
According to Eq. (\ref{newmain}) the expectation value dynamics of the observable $A$ is proportional to its auto-correlation function for
exceedingly many initial-states.
While this statement cannot be true for all initial states, we stress within this Section that it holds for huge classes of
physically relevant initial states.

In the main text the validity of (\ref{assumption}) has been exemplarily proven for initial states $\rho \propto A^N$ with $N \in {2,3}$.
In the first part of this Section we extend this proof to arbitrary but fixed $N$ at large $d$. This also ensures the vality of (\ref{assumption}) for
initial states that are analytical functions of $A$, e.g. $\exp(\lambda A), r(A)$ for sufficiently large $d$.

In the second part we numerically address the range of validity by checking the expectation-value dynamics of all eigenstates of $A$.

\subsection{Evaluation of $\Tr{{A}(t) {A}^N}$ for arbitrary but fixed 
 $N$ at large $d$}
 
Before embarking on a concrete estimate of $\Tr{{A}(t) {A}^N}$, the maximum $N$ for which Eq. (\ref{assumption}) needs to be established should be settled.
This maximum $N$ is $N=d$. To justify this, consider the following reformulation of  Eq. (\ref{assumption}):   
\begin{equation} \label{refor}
\sum_{n = 1}^d \langle a_n| A(t)| a_n \rangle a_n^N \propto \Tr{A(t) A}
\end{equation}
If this holds for $N \leq d$ it must  hold for all $\langle a_n| A(t)| a_n \rangle$ individually i.e. 
\begin{equation} \label{indi}
\langle a_n| A(t)| a_n \rangle \propto \Tr{A(t) A}
\end{equation}
since the $ a_n^N$ form a set of linearly independent functions of $n$. Having established this we are set to work towards Eq. (\ref{assumption})

To begin with, consider 
\begin{equation} \label{fab0}
 \tilde{f}^N_{ab} := A_{a b} \cdot \sum_{{i_1},...i_{N-1}}
 ( A_{b {i_1}} A_{{i_1}{i_2}} \cdots  A_{i_{N-2}i_{N-1}} A_{i_{N-1} a})\ ,
\end{equation}
where the addends are products of $N$ matrix elements $A_{ij}$. Obviously $ \tilde{f}_{ab}^N$ is a Fourier component of $ \Tr{A(t) A^N}$ with time dependence proportional 
to $\exp{-i (\epsilon_b-\epsilon_a)t}$. Since $A$ is essentially a random matrix (see Eq. (\ref{ETH})), most addends in Eq.\ \eqref{fab} are products of independent random numbers.
As such they will be real random numbers themselves, with zero mean. Hence, to an accuracy set by the law of large numbers, 
these addends will sum up to zero. However, there are index 
combinations for which the respective addends are not just products of independent random numbers  but necessarily real and positive.
(These are also the only addends that would 
''survive`` an averaging of Eq.\ \eqref{fab} over concrete implementations of $A$ as may be inferred from Isserlis theorem \cite{isserlis18}) 
First  we focus exclusively on these addends to find the ''systematic part``  of $\Tr{A(t) A^N}$. 
We come back to the ''random`` or ''fluctuating`` part below.

An index combination yields a sure positive, systematic contribution if and only if each individual matrix element  $A_{ij}$  in  Eq.\ \eqref{fab0} 
appears for an even number of times. Since for even $N$ there is an odd number of matrix elements in Eq.\ \eqref{fab0}, 
the systematic contribution vanishes in this case. This already establishes  Eq. (\ref{assumption}) for even $N$. For $N$ odd
there are very many index combinations for which each  individual matrix element appears for an even number of times. 
Consider first the two following types of such index combinations
\begin{eqnarray} \label{fab}
 \tilde{\alpha}^N_{ab} &:=& A_{a b} \cdot \sum_{{i_1},...i_{(N-1)/2}} 
 ( A_{b {i_1}} A_{{i_1}{i_2}} \cdots  A_{i_{1}{b}} A_{b a}),\ \\
\tilde{\beta}^N_{ab} &:=& A_{a b} \cdot \sum_{{i_1},...i_{(N-1)/2}} 
 ( A_{b {a}} A_{{a}{i_1}} \cdots  A_{i_{2}   i_{1}} A_{{i_1} a}),\  
\end{eqnarray}
These two contributions to $\tilde{f}^N_{ab}$ feature the maximum number of ''free indices`` i.e., indices that are summed over, under the condition that 
each matrix element has to 
appear at least twice. This number is $(N-1)/2$. There are much more sure positive index combinations, however, they all have  at most $(N-3)/2$ free indices.
 The number of free indices is crucial since each free
index gives rise to a multiplicity on the order of $d$ to the respective contribution due to the corresponding summation. The number of index combinations
 that lead to 
sure positive 
index combinations with less than $(N-1)/2$ free indices depends on $N$. Their number grows (rapidly) with $N$. However, at any fixed $N$ the 
contributions 
$\tilde{\alpha}^N_{ab}, \tilde{\beta}^N_{ab}$ will become more and more dominant with larger dimension $d$. Above some $d$ we may thus approximate 
\begin{equation} 
 \label{domcon}
  \tilde{f}^N_{ab} \approx \tilde{\alpha}^N_{ab} +  \tilde{\beta}^N_{ab}.
\end{equation}
While it is not obvious if this approximation is justified up to  $N=d$, we focus on cases where Eq. \ref{domcon} is valid in the paper at hand. 
A more thorough analysis of the $N\approx d$ case, 
will be the subject of a following publication. Taking Eq. (\ref{domcon}) for granted we obtain
\begin{eqnarray}
	\tilde{f}^N_{ab} &=&  A^2_{a b} \cdot  (P_b^N +P_a^N) \label{diff1}\\
	P_{b}^N &:=& \sum_{{i_1},...i_{(N-1)/2}} 
	( A_{b {i_1}} A_{{i_1}{i_2}} \cdots  A_{i_{1}{b}} ) \nonumber
\end{eqnarray}
$P_{b}^N$ may be interpreted as a sum over certain paths on the set of the indices where each path features a corresponding weight:
 Each path has to start at $b$, 
it has to end at $b$ and it must take each transition for an even number of times, i.e, at least twice. The total number of transitions is $N-1$.
 The number of different indices through which such a path ventures is the number of free indices, thus the paths with the largest  number of different 
 indices feature $(N-1)/2$ free indices, in accord with the above statement. The weight of each path is the product of all the squares $A_{ij}^2$ of the
 matrix elements corresponding to the transitions through which is went. Calculating (an estimate) of $P_{b}^N$ is an ambitious endeavor, closely related 
 to the derivation of Wigner's "semi-circle law" for the spectra of random matrices. Fortunately there is no need to do this here. The following
 two observations suffice: i) The statistical properties of the matrix $A$ do not depend on individual indices, they only depend on the 
 differences $\epsilon_i-\epsilon_j$. 
 Hence, up to a (small) statistical error, $P_{b}^N$ cannot depend on $b$ (up to finite size effects, cf. below). Much like the return probability of a 
 particle in a 
 disordered but homogenous medium does
 not depend on the starting point. ii) For each path that ventures through the transition  $a \leftrightarrow b$ two, four, etc. times, there are (at least) 
 $\propto d$ paths 
 that do not do so. Thus at large $d$, these paths have negligible weight. As a consequence $\tilde{f}^N_{ab}$ has a dominant contribution proportional to
 $ A^2_{a b}$ and only negligible contributions proportional to $ A^4_{a b},A^6_{a b} $, etc. While the first observation is strictly correct in the limit of 
 $\omega_\text{max}/E \rightarrow 0$, it is not strictly  correct outside this limit. If $b$ is close to one of the edges, i.e., $\epsilon_b \approx \pm E$,
 the paths become affected by the vicinity  to the edge. Much like the above return probability may be different if the starting point is sufficiently 
 close to an edge of the disordered medium. Here we assume, however that the resulting dependence of $P_{b}^N$ is such that $P_{b}^N= P^N(\epsilon_b) $ 
 does not change much on the scale of $\omega_\text{max}$.
Equipped with these observations we now return to Eq. (\ref{diff1}).
Recalling that $\tilde{f}^N_{ab}$ are the Fourier components of the 
respective correlation functions yields:
\begin{align}
\begin{split}
 \label{putog}
 \Tr{A(t) A^N}&_{\text{sur. pos.}} \\
	 \propto \sum_{a,b} & A^2_{a b} \exp{-i(\epsilon_a -\epsilon_b)t}(P_b^N +P_a^N).
\end{split}
\end{align}
Employing the index transformation
\begin{equation} 
 \label{transform}
 \overline{\epsilon}:=\frac{\epsilon_a +\epsilon_b}{2}, \quad \omega:=\epsilon_a -\epsilon_b
\end{equation}
Eq. (\ref{putog}) may be rewritten as:
\begin{align}
\begin{split}
 \label{putog1}
 \Tr{A(t) A^N}&_{\text{sur. pos.}} \\ 
     \propto \sum_{\overline{\epsilon},\omega} & A^2_{\overline{\epsilon} \omega} 
 \exp{-i\omega t}(P_{\overline{\epsilon}+\omega/2}^N +P_{\overline{\epsilon}-\omega/2}^N).
 \end{split}
 \end{align}
We proceed by exploiting two facts: i) the $ A^2_{\overline{\epsilon} \omega}$ and the $P_{\overline{\epsilon}\pm \omega/2}^N$ are (approximately) uncorrelated.
ii) the $A^2_{\overline{\epsilon} \omega}$ depend on $\overline{\epsilon}$ only statistically, the systematic dependence is only on $\omega$, cf. Eq. (\ref{ETH}). 
Exploiting these facts allows to recast Eq. (\ref{putog1}) as 
\begin{align}
\begin{split}
 \label{putog2}
 \Tr{A(t) A^N}_{\text{sur. pos.}} \propto & \sum_{\omega} A^2_{\overline{\epsilon} \omega} \exp{-i\omega t} \\
 & \sum_{\overline{\epsilon}}(P_{\overline{\epsilon}+\omega/2}^N +P_{\overline{\epsilon}-\omega/2}^N).
\end{split}
\end{align}
The second sum over $\overline{\epsilon}$ is approximately independent of $\omega (\leq \omega_{\text{max}})$  unless a substantial fraction of the weight of the 
function $P_a^N(0)$ is concentrated within  a range of width $\omega_{\text{max}}$ at the edges $-E, E$. Following the above observation i), however, 
this is not to be expected. Hence  Eq. (\ref{putog1}) may again be rewritten as 
\begin{equation} 
 \label{putog2}
 \Tr{A(t) A^N}_{\text{sur. pos.}} \propto \sum_{\overline{\epsilon},\omega} A^2_{\overline{\epsilon} \omega}. 
 \exp{-i\omega t}
\end{equation}
Realizing that the right hand side is just the Fourier transform of $\Tr{A(t) A}$ this yields 
\begin{equation} 
 \label{putog3}
 \Tr{A(t) A^N}_{\text{sur. pos.}} \propto    \Tr{A(t) A}
\end{equation}
and thus completes the justification of Eq. (\ref{assumption}).

In the remainder we analyze the influence of the non-sure positive contribution which mainly give rise to deviations/fluctuations , i.e., the ''$ \approx $`` relation
in Eq. (\ref{assumption}). We aim at estimating the scaling of these (squared) deviations with the dimension $d$. To this end we define $x_N(t)$:
\begin{equation} 
 \label{devn1}
 x_N(t) :=\frac{1}{d}( \Tr{A(t) A^N} - \Tr{A(t) A^N}_{\text{sur. pos.}})  
\end{equation}
Note that the prefactor of $d^{-1}$ renders $d^{-1}  \Tr{A(t) A^N}$ itself independent of $d$ in the limit of large $d$. Again, 
we are eventually interested in the  sure positive contributions to $|x_N(t)|^2$, denoted as $|x_N(t)|^2_{\text{sur. pos.}}$. The other contributions are expected to 
be of vanishing impact in the limit of large $d$. Writing $|x_N(t)|^2$ out explicitly yields
\begin{eqnarray} 
 \label{devn2}
|x_N(t)|^2 &= & \frac{1}{d^2}\sum_{\text{indeces} \setminus  \alpha, \beta} A_{ab} A_{b i_1}\cdots A_{i_{N-2}a } \nonumber \\
 &\cdot&   A_{cd} A_{d j_1}\cdots A_{j_{N-2}c } \nonumber \\
&\cdot& \exp{-i(\epsilon_a - \epsilon_b - \epsilon_c + \epsilon_d)t}
\end{eqnarray}
where $\text{indices} \setminus  \alpha, \beta$ stands for :''all indices, i.e., index combinations, but without those which give rise to the 
$\tilde{\alpha}^N_{ab}, \tilde{\beta}^N_{ab}   $ in Eq. (\ref{fab})``. This somewhat involved construction ensures the subtraction of the sure positive part as defined in
Eq. (\ref{devn1}).
Again, the sure positive contributions to  $|x_N(t)|^2$ arise from index combinations for which all matrix elements appear to an even power, i.e., at least 
squared. Following the same scheme as in Eq. (\ref{fab}) we find that the index combinations with the largest number of free indices are  characterized by 
$a=c, b=d, i_n =j_n$. This yields: 
\begin{equation} 
 \label{devn3}
|x_N(t)|^2_{\text{sur. pos.}} \approx  \frac{1}{d^2}\sum_{\text{indeces} \setminus  \alpha, \beta} A^2_{ab} A^2_{b i_1}\cdots A^2_{i_{N-2}a }
\end{equation}
The index combinations that would appear in Eq. (\ref{devn3}) but are excluded by the "$\setminus  \alpha, \beta$"  are the ones for which each $A_{ij}^2$ 
appears at least twice (such as to form 
$A_{ij}^4$). Consequently  the number of free indices corresponding to those excluded index  combinations is of order $N/2$ 
while the number of the free indices of  the 
"non-excluded" index combinations in the summation of Eq. (\ref{devn3}) is of order $N$. Thus, creating only a negligible error, 
we may drop the exclusion of said index combinations, obtaining 
\begin{equation} 
 \label{devn4}
 |x_N(t)|^2_{\text{sur. pos.}} \approx  \frac{1}{d^2}\sum_{\text{indeces}} A^2_{ab} A^2_{b i_1}\cdots A^2_{i_{N-2}a }
\end{equation}
As, according to Eq. (\ref{ETH}) the $N$ matrix elements scale as $A_{ij}^2 \propto d^{-1}$ and there are $N$ summations of $d$ indices in Eq. (\ref{devn4}) we eventually find 
\begin{equation} 
 \label{devn5}
 |x_N(t)|^2_{\text{sur. pos.}} {\propto} \frac{1}{d^2}
\end{equation}

While this result could in principle be compared to numerics directly, we resort here to a check of consistency of our much more
 detailed numerical findings with 
the result in Eq. (\ref{devn5}). Rather than addressing the $ \Tr{A(t) A^N}$ (for limited $N$) we numerically analyze dynamics of the form 
$\bra{a_j}A(t)\ket{a_j}$. These data are more detailed in the sense that $ \Tr{A(t) A^N}$ may conveniently be computed from the set of all 
$\bra{a_j}A(t)\ket{a_j}$. This way the consistency will eventually be demonstrated. 
We start, however, by postulating  a specific form of the $\bra{a_j}A(t)\ket{a_j}$ which
is suggested by the numerical findings, cf. Fig. \ref{error_dist}:
\begin{equation} 
 \label{devei1}
 \bra{a_j}A(t)\ket{a_j} \stackrel{!}{=} a_j f(t) + \frac{g_j (t)}{\sqrt{d}} Y_j \ ,
\end{equation}
where $Y_j$ are independent random Gaussian numbers with zero mean and unit variance. $g_j(t)$ is a function that varies very mildly with $j$.
Recalling Eq. (\ref{refor}) and identifying $\sum_j a_j^{N+1}f(t) =  \Tr{A(t) A^N}_{\text{sur. pos.}}$ 
we re-express $x_N(t)$ based on Eq. (\ref{devei1}):
\begin{equation} 
 \label{devei2}
x_N(t) = \frac{1}{d}\sum_j a_j^N  \frac{g_j (t)}{\sqrt{d}} Y_j
\end{equation}
Exploiting Eq. (\ref{devei1}) it is straightforward to compute
\begin{equation} 
 \label{devei3}
\langle |x_N(t)|^2 \rangle = \frac{1}{d^2}\sum_j a_j^{2N}  \frac{g^2_j (t)}{{d}} \propto \frac{1}{d^2}.
\end{equation}
Comparing this result to Eq. (\ref{devn5}) completes the demonstration of consistency and strongly supports the conjecture
 implemented by Eq. (\ref{devei1})

\subsection{Numerical checkup of (\ref{newmain}) for Eigenstates of $A$}
We now turn to the numerical check of the expectation-value dynamics of the eigenstates of the observable.
We generated the observable in such a way that its auto-correlation function $\Tr{A(t)A}$ is proportional to $g(t)$.
For our numerical investigations we chose the reference-function redrawing the contours of the Burj-Kalifa ($g(t)$).
Some dynamics are shown in Fig. \ref{dynamics}. 
The expectation-value dynamics of each eigenstate $\ket{a_j}$ of $A$ appear to be very similar to this auto-correlation function.
To quantify the error we define the deviation at time $t$ in the following way:
\begin{equation}
 \tensor*[]{D}{_{a_j}^t} = \bra{a_j} A(t) \ket{a_j} - a_j \frac{\Tr{A A(t)}}{\Tr{A^2}}
\end{equation}
Formally this quantity is very similar to $\frac{g_j(t)}{\sqrt{d}} Y_j$ defined in (\ref{devei1}), but while the properties of $\frac{g_j(t)}{\sqrt{d}} Y_j$
are simply conjectured, $\tensor*[]{D}{_{a_j}^t}$ refers to numerical data.
One aim in the following analysis is to show that the actual distribution of errors is compatible with the statistical properties of $\frac{g_j(t)}{\sqrt{d}} Y_j$.

We start by checking the dependence of the errors on the index of the eigenstate $j$. 

We therefor fix $t = 10$ and $d={7000, 50000}$ and plot the errors as a function of the eigenvalue $a_j$ of the initial state $\rho_0 = \ket{a_j}\bra{a_j}$ (Fig. \ref{error_dist}).

%The errors for fixed $t=10$, $j$ and $d={7000, 50000}$ appear to be normal distributed among multiple instances with different random numbers.
%While the mean and the variances slitely depend on the position in time and in the spectrum. 

Fig. \ref{error_dist} indicates that there is no systematic dependence of the error on the position in the spectrum.
Moreover the errors appear to be normal distributed.
The numerics clearly show that the absolute errors decrease with dimension $d$.
This dependence on $d$ is studied in more detail in a finite-size scaling at the end of this section.

Up to now we focused on a single point in time.
To drop this random choice, we define two new error-measures: $\tensor[]{D}{_{a_j}}$, which quantifies the
squared deviation of a dynamics averaged over time and $\tensor[]{D}{_d}$, which is the squared deviation averaged over time and over all
eigenstates of $A$:
\begin{equation}
 \tensor*[]{D}{_{a_j}} = \frac{1}{M} \sum_{m=0}^{M-1} \left( \tensor*[]{D}{^{m \cdot \Delta t}_{a_j}} \right) ^2
\end{equation}

\begin{equation}
 \tensor*[]{D}{_d} = \frac{1}{d} \sum_{j=1}^d \tensor*[]{D}{_{a_j}}
\end{equation}

$\Delta t = 0.25$ and $M=200$ denote the (numerical) time-step and the number of steps in time, respectively.
%Please note that the latter definition $\tensor*[]{D}{_d}$ is the time-discrete form of $D(d)$ (\ref{mdev}).

\begin{figure}
\centering
\includegraphics[width=0.48\textwidth]{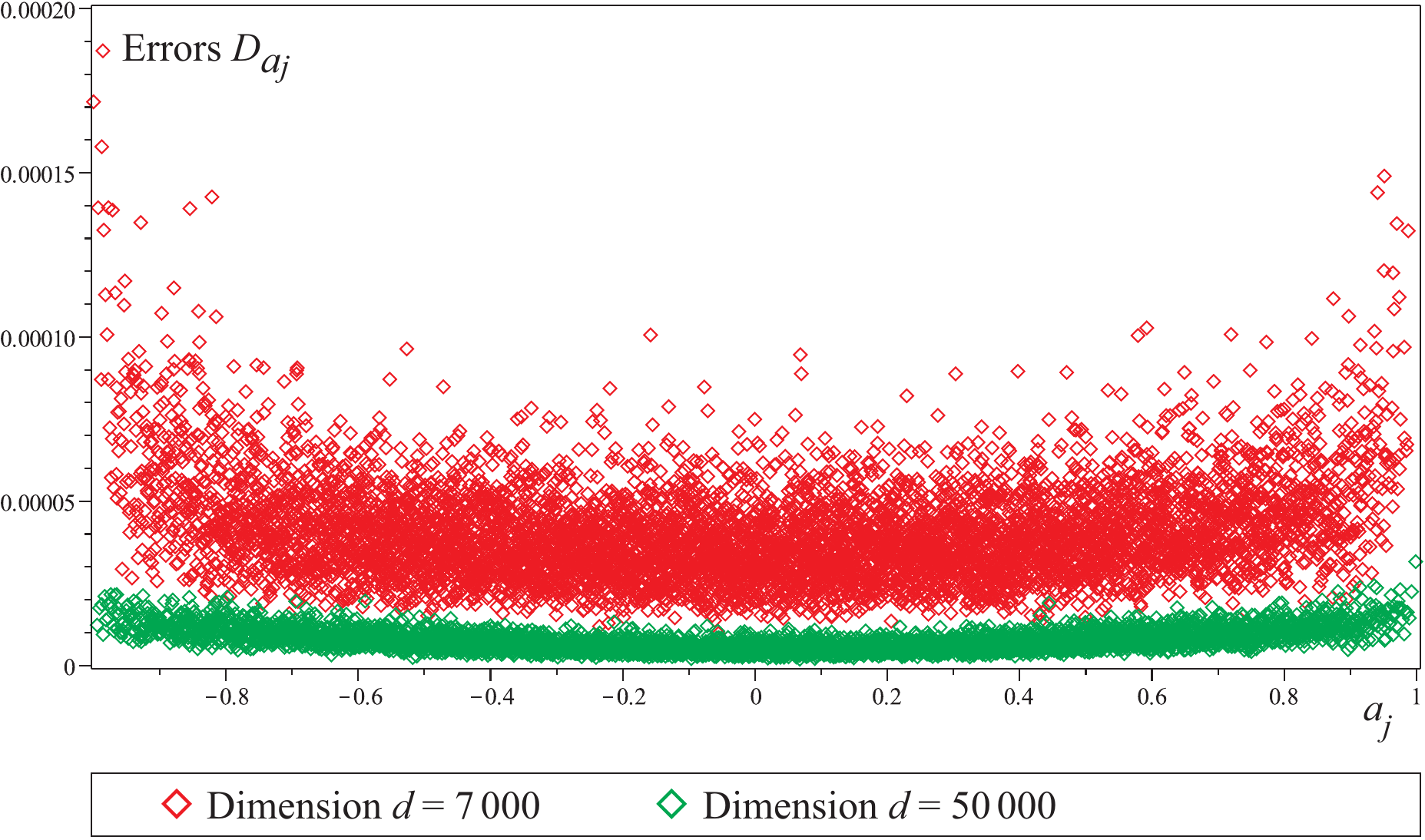} 
\caption{
	The time-averaged quadratic error $\tensor*[]{D}{_{a_j}}$ only marginally depends on the eigenstate $\rho_0=\ket{a_j}\bra{a_j}$, which serves as the initial state.
	Furthermore the deviations from the reference-function in the larger system $d=50000$ are significantly smaller than the corresponding deviations for $d=7000$.
}
	\label{error_by_eigenvalue}
\end{figure}

Fig. \ref{error_by_eigenvalue} shows the averaged deviations of the expectation-value dynamics from the reference function for all eigen-states of $A$.
The errors appear to be only marginally dependent on the position in the spectrum of $A$.

Fig. \ref{error_by_eigenvalue} furthermore indicates that the deviations from the reference-function decrease for larger dimension $d$.
To address the dependence of the errors on the system size $d$ we plot the averaged errors $\tensor*[]{D}{_d}$ as a function of the reciprocal dimension $1/d$ (Fig. \ref{Richter_Errors_finite_size}).

This finite-size scaling indicates that the mean quadratic error $\tensor*[]{D}{_d}$ is proportional $\frac{1}{d}$.
This suggests that the variances of the errors $\tensor*[]{D}{_{a_j}^t}$ each are proportional to $\frac{1}{d}$.

Thus the properties of the error distribution are in accord with the assumptions made in (\ref{devei1}), which in turn implies
the $\frac{1}{d^2}$-error scaling for $N \ll d$ found in (\ref{devei3}).

\section{Typicality} \label{app_typicality}
In this Section a derivation  of Eq. (\ref{approx}) is presented. It is similar to a comparable analysis in Ref. \cite{reimanngemmer}. 
Consider a ensemble of pure states given by
\begin{equation}
\ket{\psi} =\bra{\phi}\ket{\phi}^{-1/2} \ket{\phi}, \quad 
\ket{\phi} = \sqrt{r(A)}\ket{\xi}
\label{offdiaapp}
\end{equation}
where $r(A)$ is a nonnegative, ``smooth'' function, i.e., $|r(a_{j+1})-r(a_{j})|/| a_{j+1}-a_{j}| < d$, with respective expectation values $\bra{\psi} A(t) \ket{\psi}$ 
\begin{equation}
 \label{offdia1}
 \bra{\psi} A(t) \ket{\psi} =  \frac{\bra{\xi}\sqrt{r(A)}  A(t) \sqrt{r(A)} \ket{\xi}}{\bra{\xi}r(A) \ket{\xi}        }.
\end{equation}
We analyze the statistical properties of the numerator first. Let the overbar $\overline{\cdots}$ indicate the average over all $\ket{\xi}$ and $\sigma^2( \cdots  )$ the respective variance. 
Following \cite{lloyd, gemmer} we obtain: 
\begin{eqnarray}
&& \overline{\bra{\xi}\sqrt{r(A)}  A(t) \sqrt{r(A) }  \ket{\xi}  } = \mathbb{E}[  r(A)  A(t)]  ,   
 \label{nume1} \\ 
 &&\sigma^2( \bra{\xi}\sqrt{r(A)}  A(t) \sqrt{r(A) }  \ket{\xi}  )   =  \\
 &&\frac{\chi^2[\sqrt{r(A)}  A(t) \sqrt{r(A)}]}{d+1} \leq 
 \frac{\mathbb{E}[ r(A)A^2] }{d+1}      \label{nume2}       
\end{eqnarray}
 where $\mathbb{E}[\cdots]$ denotes the mean and $\chi^2[\cdots]$ the variance of the spectrum of the respective operator. As $\mathbb{E}[ r(A)A^2]$ converges against
 a fixed value for large $d$ and smooth $r(A)$, the variance $\sigma^2( \bra{\xi}\sqrt{r(A)}  A(t) \sqrt{r(A) }  \ket{\xi}  ) $ has an upper bound that essentially scales as $d^{-1}$. Hence
 \begin{equation}
 \label{numeapp}
 \bra{\xi}\sqrt{r(A)}  A(t) \sqrt{r(A) }  \ket{\xi}   \approx \mathbb{E}[  r(A)  A(t)] = \frac{ \Tr{r(A)  A(t)}}{ d}
\end{equation}
is a very good approximation for all $\ket{\xi}$ except for a fraction of size of at most $\propto d^{-1/2}$. We now perform an analogous analysis for the denominator of 
Eq. (\ref{offdia1}):
\begin{equation}
 \label{denom}
 \overline{  \bra{\xi}r(A) \ket{\xi} }  =  \mathbb{E}[r(A)], \quad   \sigma^2( \bra{\xi}f(A) \ket{\xi} )  =  \frac{ \chi^2[r(A)]}{(d+1)} 
\end{equation}
 As $ \chi^2[r(A)]$ converges against
 a fixed value for large $d$ and smooth $r(A)$, the variance $\sigma^2( \bra{\xi}f(A) \ket{\xi} ) $ essentially scales as $d^{-1}$. Hence
 \begin{equation}
 \label{denomeapp}
 \bra{\xi}r(A) \ket{\xi}   \approx \mathbb{E}[r(A)] = \frac{ \Tr{r(A)} }{ d}
\end{equation}
is a very good approximation for all $\ket{\xi}$ except for a fraction of size of  $\propto d^{-1/2}$. Inserting Eq. (\ref{numeapp}) and  Eq. (\ref{denomeapp}) 
into Eq. (\ref{offdia1}) yields
 \begin{equation}
 \label{approxapp}
\bra{\psi} A(t) \ket{\psi} \approx  \frac{ \Tr{r(A)  A(t)}}{ \Tr{r(A)} }
\end{equation}
as a good approximation  for all $\ket{\psi}$ except for a fraction of size of at most $\propto d^{-1/2}$. This establishes Eq. (\ref{approx}).

\end{document}